\makeatletter
\declare@file@substitution{revtex4-1.cls}{revtex4-2.cls}
\makeatother
\documentclass[twocolumn,onecolappendix]{aastex631}

\usepackage{amsmath}
\usepackage{graphicx}
\usepackage{amstext}

\begin{document}

\title{Multi-wavelength Emission of Gamma-ray Burst Prompt Phase. \\I. Time-resolved and Time-integrated Polarizations}


\author{Jia-Sheng Li}
\author[0000-0001-5641-2598]{Mi-Xiang Lan}
\author{Hao-Bing Wang}
\affiliation{Center for Theoretical Physics and College of Physics, Jilin University, Changchun, 130012, China; lanmixiang@jlu.edu.cn}

\begin{abstract}
The time-integrated polarization degree (PD) at prompt optical band of gamma-ray burst (GRB) was predicted to be less than $20\%$, while the time-resolved one can reach as high as $75\%$ in photosphere model. Polarizations in optical band during GRB prompt phase had not been  studied under framework of the magnetic reconnection model. Here, a three-segment power laws of the energy spectrum is used to reconstruct the Stokes parameters of the magnetic reconnection model. The multi-wavelength light curves and polarization curves from the optical band to MeV gamma-rays in GRB prompt phase are studied. We found depending mainly on the jet dynamics there is a long lasting high PD phase at all calculated energy bands for the typical parameter sets. The time-resolved PD could be as high as $50\%$, while the time-integrated one is roughly $17\%$) in optical band. It can reach $60\%$ for the time-resolved PD in X-rays and the time-integrated one is around $(30-40)\%$. The polarization angle (PA) evolution is random in both optical and gamma-ray bands for the photosphere model, while it is roughly a constant in the synchrotron models. Therefore, future time-resolved PA observations in the prompt optical or gamma-ray band could distinguish between the photosphere and the synchrotron models.

\end{abstract}


\section{Introduction}

Gamma-ray bursts (GRBs) are sudden intensification of gamma rays in the universe. Even if GRB was first detected in 1967, there are still many mysteries to solve after decade years of studies. They are divided into long and short bursts depending on whether or not the duration of the prompt-emission phase exceeds 2 s. Long bursts usually result from the collapse of a massive star \citep{1993ApJ...405..273W,1999Natur.401..453B,2001ApJ...550..410M,2003ApJ...599L..95M,2003Natur.423..847H}, while the short bursts are from the merger of the two compact objects \citep{1992ApJ...395L..83N,2017ApJ...848L..13A,2017ApJ...848L..14G,2018PhRvL.120x1103L}. For a black hole central engine, Blandford–Znajek mechanism \citep{1977MNRAS.179..433B} would work and the magnetic field configuration (MFC) in the ejecta will be toroidal, while for a magnetar, a stripped wind would be blew out and the MFC in the wind is aligned \citep{2001A&A...369..694S}. It was pointed out that the evolution patterns of the polarization angle (PA) will be different for an aligned and a toroidal field in the ejecta during early optical afterglow, so the PA evolution patterns can be used as a probe of the GRB central engine \citep{2016ApJ...816...73L}.

In order to explain the observations of the GRB prompt phase, the internal shock model \citep{1992ApJ...395L..83N,1994ApJ...430L..93R} and the photospheric model \citep{1994MNRAS.270..480T,2000ApJ...529..146E,2000ApJ...530..292M,2005ApJ...628..847R,2009ApJ...700L..47L,2011ApJ...737...68B,2011ApJ...732...49P,2011ApJ...732...26M,2011ApJ...731...80N,2012ApJ...746...49X,2013ApJ...772...11R,2013ApJ...767..139B,2013MNRAS.428.2430L,2013ApJ...765..103L} were proposed. In the last decade, the magnetic reconnection model \citep{2008A&A...480..305G,2011ApJ...726...90Z,2016MNRAS.459.3635B,2016ApJ...816L..20G} also becomes popular. It assumes that the highly magnetized central engine ejects highly magnetized shells and these shells will collide with each other, leading to the magnetic reconnection process \citep{2011ApJ...726...90Z}. Then the free magnetic energy is released to accelerate the electron to produce the synchrotron radiation. 

\cite{2020ApJ...892..141L} had studied the polarization of the GRB prompt phase under the framework of the magnetic reconnection model. The predicted upper limit of the polarization degree (PD) of the magnetic reconnection model is $\sim50\%$ around gamma-ray band \citep{SL2024}, which is different from a roughly zero polarization in the photosphere model \citep{2018ApJ...856..145L}. The polarization studies of the above models mainly focused on the energy band above soft X-ray. Then it is extended to the optical band in the photosphere model \citep{2022ApJ...926..104P}. The results showed that the time-integrated PD in the optical band for the photosphere model is below $20\%$, which is higher than the predicted zero polarization in gamma-ray band beyond MeV. However, the optical polarization in GRB prompt phase had not been predicted under the framework of the magnetic reconnection model.

The spectrum of the GRB prompt emission is usually fitted by the empirical formulas, i.e., the Band function \citep{1993ApJ...413..281B}, which is a two-segment power laws connected at $E_p$ (the peak of the ${\nu}F_{\nu}$ spectrum). Fitting the data from the Fermi Gamma-Ray Burst Monitor (Fermi-GBM), the low-energy photon spectral index of the Band function above energy bands of hard X-ray is indicated to be concentrated around -1 \citep{2016A&A...588A.135Y,2021ApJ...913...60P}, which is inconsistent with the predicted value of -3/2 for the fast-cooling synchrotron emission. Then theoretically \cite{2014NatPh..10..351U} had obtained the value of $-1$ for the low-energy photon spectral index in a relatively weak decaying magnetic field and it approaches -3/2 when the magnetic field becomes stronger. In a wider energy band down to roughly soft X-ray even to optical band, some studies \citep{2017ApJ...846..137O,2018A&A...616A.138O,2019A&A...628A..59O,2019A&A...625A..60R,2021A&A...652A.123T} had suggested that there might be a low-energy break in the GRB spectrum and the break energy ($E_b$) is around a few keV. They found the photon spectral indices are -2/3 below $E_b$, -3/2 between $E_b$ and $E_p$, and -2.3 above the $E_p$, respectively. And their results support the fast-cooling synchrotron emission as the main radiation mechanism in the GRB prompt phase.

\cite{2020ApJ...892..141L} used the Band function \citep{1993ApJ...413..281B} to construct the Stokes Parameters to study the polarizations of the $\gamma$-rays band in GRB prompt phase. In this paper, a three-segment power-law spectrum is used to rebuild the Stokes Parameters (in Section \ref{sec:The Models}), and the observational energy band can be extended down to the lower optical band. Here, the multi-wavelength (from optical to MeV $\gamma$-rays) light curves, polarization curves and time-integrated polarizations in the GRB prompt phase are predicted, and also the influences of the key parameters on the time-resolved and time-integrated polarizations are discussed in Section \ref{sec:Numerical Results}. Finally, our conclusions and discussion are arranged in Section \ref{sec:Conclusions and Discussion}.

\section{The Models}\label{sec:The Models}

According to the former studies \citep{2015ApJ...808...33U,2016ApJ...825...97U,2018ApJ...869..100U}, a relativistic jet shell expands along the radial direction from the central engine at redshift $z$. The electrons are isotropically injected into the shell from the radius $r_{on}$ to $r_{off}$  and emit synchrotron photons in the magnetic field. The emitted photons are also assumed to be isotropic in the comoving frame of the jet shell.

Under framework of the magnetic reconnection model, depending on different MFCs in the radiation region, the jet shell could accelerate (corresponding to an aligned field) or keep as a roughly constant velocity (corresponding to a toroidal field) during GRB phase \citep{2002A&A...387..714D}. The bulk Lorentz factor $\Gamma$ would roughly evolve as a power law with radius \citep{2002A&A...387..714D}. It reads
\begin{equation}
\Gamma(r)=\Gamma_0(r/r_0)^s,
\end{equation}
where $r_0$ is the normalization radius. It is predicted that $s=0.35$ for an aligned field in the emitting region \citep{2002A&A...387..714D}, which is the parallel latitude circles in the jet surface \citep{2001A&A...369..694S,2019ApJ...870...96L}. While a toroidal filed with $s=0$ is the concentric circles around the jet axis in the jet surface.

In the following, the parameters with superscript “ $'$ ” indicate the comoving-frame quantities. The spectral power of the emission from single electron can be expressed as
\begin{equation}
P'_{\nu'}(\nu')=P'_0H_{en}(\nu'),
\end{equation}
where $P'_0$ shows the magnitude and $H_{en}(\nu')$ gives the shape of the photon spectrum. The comoving frequnecy $\nu'$ is $\nu'=\nu(1+z)/\mathcal{D}$. And $\nu$ is the observational frequency and $\mathcal{D}=1/\Gamma(1-\beta_V\cos\theta)$ is the Doppler factor. $\beta_V$ is dimensionless velocity of the shell. And $\theta$ is the angle between the velocity of the jet element and the line of sight. 

The $P'_0$ reads \citep{1979rpa..book.....R}
\begin{equation}
P'_0=\frac{3\sqrt{3}}{32}\frac{m_ec^2\sigma_TB'}{q_e},
\end{equation}
where $m_e$ and $q_e$ are the mass and charge of the electron, respectively. $c$ is the speed of light and $B'$ represents the strength of the magnetic field.
\begin{equation}
B'(r)=B'_0(r/r_0)^{-b},
\end{equation}
where $b$ is the decay index. Due to the expansion of the jet shell and the magnetic reconnection process, the decay index would be larger than 1 \citep{2002A&A...387..714D}.

For the $H_{en}(\nu')$, we assume a three-segment-power-laws spectrum for the single-energy electron to mimic the energy spectrum of the radiation from the power-law distributed electrons.
\begin{equation}\label{Hen}
H_{en}(\nu')=\begin{cases}
(\nu'/\nu'_1)^{\alpha_1+1}, & \text{$\nu'< \nu'_1$}, \\ (\nu'/\nu'_1)^{\alpha_2+1}, & \text{$\nu'_1< \nu'< \nu'_2$}, \\ (\nu'_2/\nu'_1)^{\alpha_2+1}(\nu'/\nu'_2)^{\beta+1}, & \text{$\nu'> \nu'_2$},
\end{cases}
\end{equation}
where $\alpha_1$, $\alpha_2$ and $\beta$ are the low-energy, mid-energy and high-energy photon spectral indices, respectively. The $\nu'_1$ and $\nu'_2$ are $\min(\nu'_{cool}, \nu'_{min})$ and $\max(\nu'_{cool}, \nu'_{min})$, respectively. Then $\nu'_{cool}$ and $\nu'_{min}$ read:
\begin{equation}
\nu'_{cool}=\frac{q_eB'\gamma_{cool}^2\sin\theta'_B}{2\pi m_ec},\ \ \ \ \nu'_{min}=\frac{q_eB'\gamma_{ch}^2\sin\theta'_B}{2\pi m_ec},
\end{equation}
where $\gamma_{ch}$ and $\gamma_{cool}$ are the characteristic Lorentz factor of electrons. The $\theta'_B$ is the pitch angle of the electrons in the magnetic field.

In \cite{2018ApJ...869..100U}, five variation patterns of $\gamma_{ch}$ with radius were discussed. After our calculation, there are roughly two evolution profiles of the PD curves for these five $\gamma_{ch}$ patterns, which correspond to the two peak-energy evolution modes (i.e., the hard-to-soft mode and the intensity-tracking mode). For the $i$ model with hard-to-soft mode, its $\gamma_{ch}$ can be expressed by
\begin{equation}\label{gammai}
\gamma_{ch}(r)=\gamma_{ch}^i(r/r_0)^g,
\end{equation}
where we take $g=-0.2$ \cite{2018ApJ...869..100U}. For the $m$ model with the intensity-tracking mode, its $\gamma_{ch}$ reads
\begin{equation}\label{gammam}
\gamma_{ch}(r)=\gamma_{ch}^m\times\begin{cases}
(r/r_m)^g, & \text{$r\leq r_m$}, \\ (r/r_m)^{-g}, & \text{$r\geq r_m$},
\end{cases}
\end{equation}
where $r_m$ is normalization radius and we take $g=1.0$ \cite{2018ApJ...869..100U}. The $\gamma_{cool}$ reads:
\begin{equation}
\gamma_{cool}=6\pi\dfrac{m_ec\Gamma}{\sigma_TB'^2t},
\end{equation}
where $t$ is the dynamical time in the burst source frame and $\sigma_T$ is the Thomson cross section.

For fast cooling (i.e. $\nu'_{cool}< \nu'_{min}$), $\alpha_1=-2/3$ and $\beta=-1-\dfrac{p}{2}$, where $p$ is the index of the true power-law injected electrons ($N(\gamma_e)\propto\gamma^{-p}_e$). And we take $p=2.6$ in this paper. Observationally, the fitting result of the Band function in the energy band above hard X-rays for the mid-energy photon spectral index is $-1$ \citep{2016A&A...588A.135Y,2021ApJ...913...60P}, while it is $-3/2$ when the energy band used for the analysis of the energy spectrum is extended to the optical band \citep{2017ApJ...846..137O,2018A&A...616A.138O,2019A&A...628A..59O,2019A&A...625A..60R,2021A&A...652A.123T}. Theoretically, in a decaying magnetic field the photon spectral index is $-1$ for a relatively weak field, while it approaches $-3/2$ when the magnetic field becomes stronger \cite{2014NatPh..10..351U}. So two mid-energy photon spectral indices ($\alpha_2=-1$ and $\alpha_2=-3/2$) are considered here. For slow cooling (i.e. $\nu'_{min}< \nu'_{cool}$), $\alpha_1=-2/3$, $\alpha_2=-1-\dfrac{p-1}{2}$ and $\beta=-1-\dfrac{p}{2}$.

The observed flux density $f_{\nu}$, the Stokes parameters $Q_{\nu}$ and $U_{\nu}$ here are slightly different from that in \cite{2020ApJ...892..141L}.
\begin{equation}\label{fv}
f_{\nu}=\frac{1+z}{4\pi D_L^2}\int\frac{\mathcal{D}^2}{\Gamma}\frac{c}{4\pi r}NP'_0H_{en}(\nu')\sin\theta'_Bd\phi dt,
\end{equation}
and
\begin{equation}\label{Qv}
Q_{\nu}=\frac{1+z}{4\pi D_L^2}\int\frac{\mathcal{D}^2}{\Gamma}\frac{c}{4\pi r}NP'_0H_{en}(\nu')\sin\theta'_B\Pi_p\cos2\chi_pd\phi dt,
\end{equation}
and
\begin{equation}\label{Uv}
U_{\nu}=\frac{1+z}{4\pi D_L^2}\int\frac{\mathcal{D}^2}{\Gamma}\frac{c}{4\pi r}NP'_0H_{en}(\nu')\sin\theta'_B\Pi_p\sin2\chi_pd\phi dt,
\end{equation}
where $D_L$ is the luminosity distance of the source and $N=\int R_{inj}dt/\Gamma$ is the isotropic total electron number in the shell. The $\phi$ is the angle in the plane of sky between the projection of the jet axis and projection of the radial direction of a local fluid element.

The $\Pi_p$ and $\chi_p$ are the local PD and PA, respectively. The $\Pi_p$ will be expressed as
\begin{equation}\label{localPD}
\Pi_p=\frac{\tilde{\alpha}}{(\tilde{\alpha}-2/3)},
\end{equation}
where $\tilde{\alpha}$ is the photon spectral index. $\tilde{\alpha}=\alpha_1$ for $\nu'< \nu'_1$, $\tilde{\alpha}=\alpha_2$ for $\nu'_1< \nu'< \nu'_2$, and $\tilde{\alpha}=\beta$ for $\nu'> \nu'_2$. Expressions of $\chi_p$ can be found in \cite{2016ApJ...816...73L}. 

The equal arrival time surface (EATS) effect is considered here. The photons, which are emitted from the radius $r$ at the burst-source time $t$, will arrive the observer at the observer time $t_{obs}$ \citep{1998ApJ...494L..49S,2016ApJ...825...97U}.
\begin{equation}
t_{obs}=\left[t-\frac{r}{c}\cos\theta-t_{on}+\frac{r_{on}}{c}\right](1+z).
\end{equation}

For an aligned field in the emission region, PD ($\Pi$) and PA ($\chi$) of the jet-shell radiation in GRB prompt phase are
\begin{equation}
\Pi=\frac{\sqrt{Q^2_\nu+U^2_\nu}}{f_\nu},
\end{equation}
and
\begin{equation}
\chi_{pre}=\frac{1}{2}\arctan\left(\frac{U_\nu}{Q_\nu}\right).
\end{equation}
According to \cite{2018ApJ...860...44L}, when $Q_{\nu}>0$, the final PA $\chi$ equals to $\chi_{pre}$; when $Q_{\nu}<0$, if $U_{\nu}>0$ then the PA is $\chi=\chi_{pre}+\pi/2$, if $U_{\nu}<0$ then the PA is $\chi=\chi_{pre}-\pi/2$. For a toroidal magnetic field in a roughly constant-velocity shell, $U_{\nu}$ is always 0 because of the axial symmetry and the PD of the jet shell is expressed as
\begin{equation}
\Pi=\frac{Q_\nu}{f_\nu}.
\end{equation}
If the PD changes its sign, it indicates that the PA is rotated by $\pi/2$.

\section{Numerical Results}\label{sec:Numerical Results}

Time-resolved and time-integrated polarizations in multi-wavelength are studied and the influences of the parameters are also investigated. The orientation of the aligned magnetic field is assumed to be $\delta=\pi/6$ with respect to the projection of the jet axis on the plane of sky. Unless otherwise specified, the values of the following parameters are fixed: $\Gamma_0=250$, the jet opening angle $\theta_j=0.1$ rad, observational angle $\theta_V=\theta_j/2=0.05$ rad, $B'_0=30$ G, $r_{on}=10^{14}$ cm, $r_{off}=3\times10^{16}$ cm, $\gamma_{ch}^i=5\times10^4$, $r_0=10^{15}$ cm, $\gamma_{ch}^m=2\times10^5$ and $r_m=2\times10^{15}$ cm \citep{Ghirlanda2018,Lloyd2019,RE2023,2018ApJ...869..100U}. 

\begin{figure*}
\centering
\includegraphics[scale=0.32]{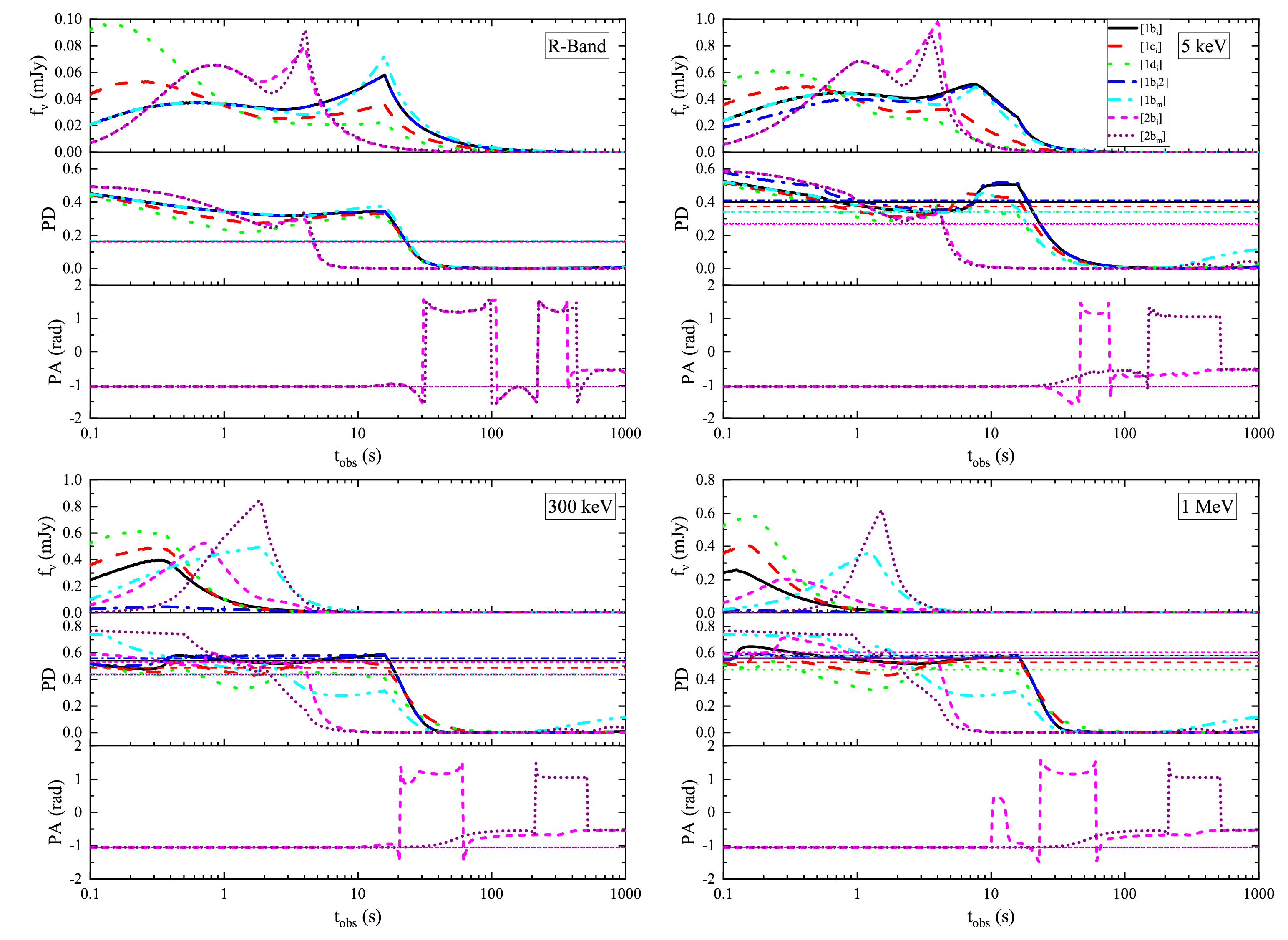}
\caption{Light curves and polarization evolutions for seven different models. Upper left, upper right, lower left, and lower right panels correspond to the observational energy bands of R-band, 5 keV, 300 keV, and 1 MeV, respectively. In each panel, top, mid and bottom panels show the light, PD and PA curves, respectively. The black solid, red dashed, green dotted, blue dash-dotted, cyan double dot-dashed, magenta short dashed, and purple short dashed lines correspond to the models of $[1b_i]$, $[1c_i]$, $[1d_i]$, $[1b_i2]$, $[1b_m]$, $[2b_i]$, and $[2b_m]$, respectively. The horizontal  reference lines show the time-integrated PDs and PAs of the corresponding model.}
\label{lc_Uhmmodel}
\end{figure*}

First, we examine the influences of the field decaying index $b$, mid-energy photon spectral index $\alpha_2$, $\gamma_{ch}$ patterns, and variation of bulk Lorentz factor $\Gamma(r)$ on the multi-band light and polarization curves. The other parameters (except the parameters with fixed values listed above) are listed in Table. \ref{Model Parameters}, and the calculation results are shown in Figure \ref{lc_Uhmmodel}.

\begin{deluxetable*}{lcccc}
\tabletypesize{\large}
\tablewidth{20pt}
\tablecaption{Model Parameters.\label{Model Parameters}}
\tablehead{ 
\colhead{Model}
&\colhead{\textcolor{black}{$b$}}
&\colhead{\textcolor{black}{$s$ (MFC)}}
&\colhead{\textcolor{black}{$\gamma_{ch}$ profile}} 
&\colhead{\textcolor{black}{$\alpha_2$}} 
}
\startdata
 $1b_i$ & \textcolor{black}{1.0} &  \textcolor{black}{0 (toroidal)}  & \textcolor{black}{Equation \ref{gammai}} & \textcolor{black}{-1.0}  \\\hline
 $1c_i$ & \textcolor{black}{1.25} &  \textcolor{black}{0 (toroidal)}  & \textcolor{black}{Equation \ref{gammai}} & \textcolor{black}{-1.0}  \\\hline
 $1d_i$ & \textcolor{black}{1.5} &  \textcolor{black}{0 (toroidal)}  & \textcolor{black}{Equation \ref{gammai}} & \textcolor{black}{-1.0}  \\\hline
 $1b_i2$ & \textcolor{black}{1.0} &  \textcolor{black}{0 (toroidal)}  & \textcolor{black}{Equation \ref{gammai}} & \textcolor{black}{-3/2}  \\\hline
 $1b_m$ & \textcolor{black}{1.0} &  \textcolor{black}{0 (toroidal)}  & \textcolor{black}{Equation \ref{gammam}} & \textcolor{black}{-1.0}  \\\hline
 $2b_i$ & \textcolor{black}{1.0} &  \textcolor{black}{0.35 (aligned)}  & \textcolor{black}{Equation \ref{gammai}} & \textcolor{black}{-1.0}  \\\hline
 $2b_m$ & \textcolor{black}{1.0} &  \textcolor{black}{0.35 (aligned)}  & \textcolor{black}{Equation \ref{gammam}} & \textcolor{black}{-1.0}  \\\hline
\enddata
\end{deluxetable*}

\begin{figure}
\centering
\includegraphics[scale=0.32]{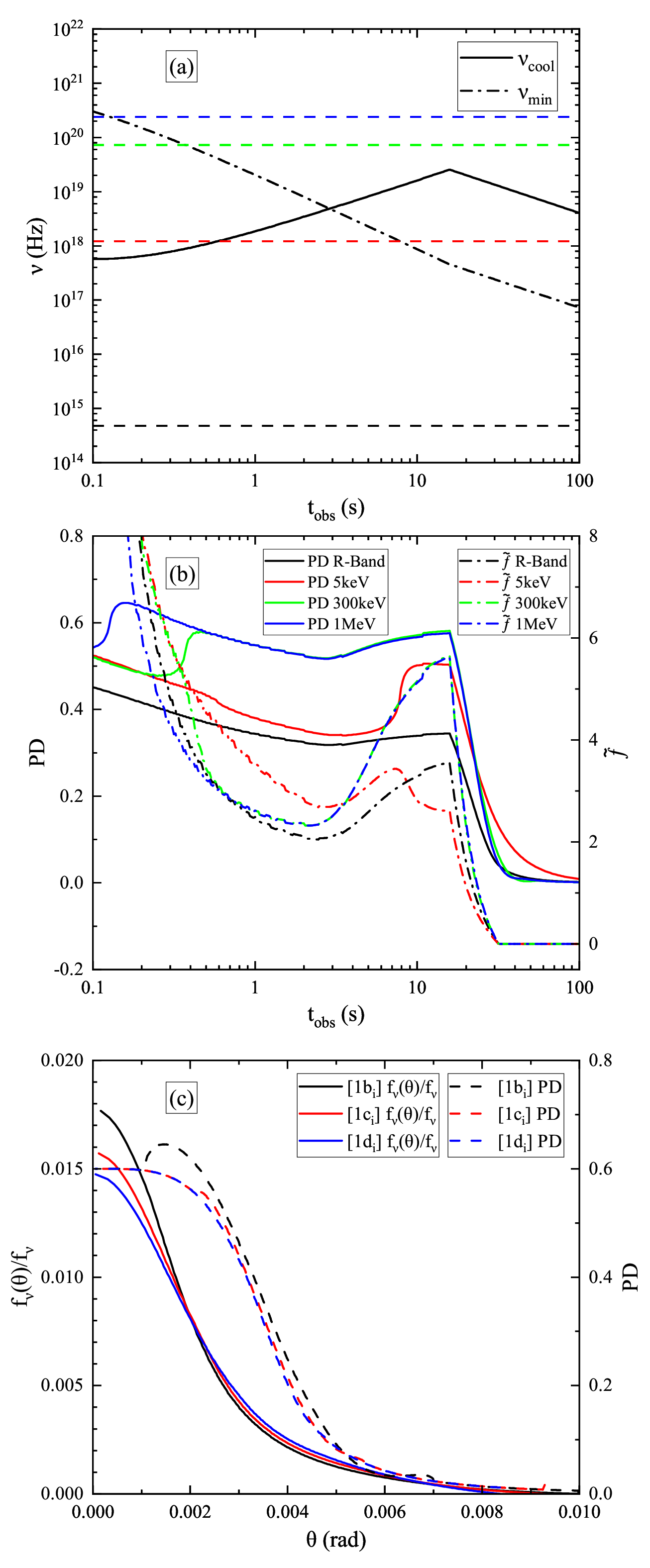}
\caption{Analysis with the $[1b_i]$ model. (a), evolutions of $\nu_{cool}$ and $\nu_{min}$ (at the radius of the maximum local flux on one EATS) with the observer time. The black-solid and dash-dot lines correspond to the $\nu_{cool}$ and the $\nu_{min}$, respectively. Black, red, green, and blue dashed horizontal reference lines correspond to the observational frequencies of R-band, 5 keV, 300 keV, and 1 MeV, respectively. (b), evolutions of the PD and $\tilde{f}$ parameter with time. The solid and dash-dot lines correspond to the PD and the $\tilde{f}$ curves, respectively. The black, red, green, and blue lines correspond to the observational energy bands of R-band, 5 keV, 300 keV, and 1 MeV, respectively. (c) The distribution of $f_{\nu}(\theta)$ (solid lines) and $PD(\theta)$ (dashed lines) with $\theta$ on the EATS of the peak time of the light curve at 300 keV. The black, red, and blue lines correspond to the models of $[1b_i]$, $[1c_i]$, and $[1d_i]$, respectively.}
\label{reason_with_tobs}
\end{figure}

For all seven models studied in Figure \ref{lc_Uhmmodel}, there are two peaks in the light curves of the optical R-band and 5 keV, while there is only one peak at 300 keV and 1 MeV. In the following, we use $[1b_i]$ model as an example to study the reasons. Because both the $\nu_{cool}$ and $\nu_{min}$ will vary with the radius, two critical frequencies at the radius of the maximum local flux on one EATS \footnote{Here, the local flux refers to the accumulated flux density at the ring between $\theta$ and $\theta+d\theta$.} are shown in Figure \ref{reason_with_tobs}. We find that the cooling is fast at early stage,and then transit to slow at late times. We define the minimum critical frequency $\nu_1\equiv$min$(\nu_{cool}$, $\nu_{min})$. The light-curve peaks usually happen when $\mid\nu-\nu_1\mid$ reaches its minimum value for a constant maximum flux $F_{\nu,max}$ \citep{1998ApJ...497L..17S}. However, the situation becomes complex for a decaying $F_{\nu,max}$. For example, in the case of 5 keV, $\nu_1(=\nu_{cool})$ equals to 5 keV at $t_{obs}=0.74$ s, and the first peak of the 5 keV light curve is just around that time. Similarly, we can see that the second peak occurs near the time when $\nu_1(=\nu_{min})$ reaches 5 keV. Therefore, for the light curves at 5 keV, the first peak is the fast-cooling peak, and the second is the slow-cooling peak. At optical R-band, the first peak is around the time when $\mid\nu_1-\nu\mid$ reaches its minimum value. Although $\nu_1$ approaches the frequency of the optical R-band at a later time, a decreasing $F_{\nu,max}$ shift the second peak to a relatively earlier time. In the higher energy bands (300 keV and 1 MeV), because of a decaying $F_{\nu,max}$ and the evolution of the $\nu_1$, there is only one light-curve peak.

It is noted that all calculated PD curves have shallowly rising stages before their steep-decay stages. To interpret this, taken $[1b_i]$ model as an example, we plot the PD curves and the evolutions of the corresponding $\tilde{f}$ parameter together as shown in Figure \ref{reason_with_tobs}. $\tilde{f}$ represents the flux ratio between $\theta\Gamma(r)<1$ and $\theta\Gamma(r)>1$ \citep{2020ApJ...892..141L}.
\begin{equation}
\tilde{f}=\frac{\int_{r_c}^{r_{max}}dF_{\nu}}{\int_{r_{min}}^{r_c}dF_{\nu}},
\end{equation}
where $r_{min}=\max(r_{on},r(t_{obs},\theta=\theta_V+\theta_j))$, $r_{max}=\min(r_{off},r(t_{obs},\theta=0))$ and $\theta\Gamma(r_c)=1$. It is found that the two quantities are positively correlated except for $\tilde{f}=0$ or $\tilde{f}=\infty$. With the increase of the $\tilde{f}$ ($\tilde{f}\neq0$ and $\tilde{f}\neq\infty$), the proportion of the radiation at high latitudes will decrease (i.e., the proportion of the radiation outside the $1/\Gamma$ cone will decrease), then the jet PD will increase and vice versa.

The early high PD phase in the PD curve would be followed by a steep decay phase at late time. Independent of the observational energy band, the begining time of the late-time steep decay phase of the PD curve is same for the same model. For on-axis observations with $\Gamma_0\theta_j>1$, PD will decay rapidly when the jet shell expands beyond $r_{off}$. Therefore, the lasting timescales of the high PD phase is determined by the dynamics of the jet shell. For the same dynamics, i.e., the same bulk-Lorentz-factor variation pattern, the lasting timescales of the high PD phase would be same. The PAs are roughly constants during the main radiation episodes for the four calculated energy band, which is different from a randomly evolving PA in the photosphere model \citep{2022ApJ...926..104P}.

The magnetic field strength with a larger $b$ value is larger at small radius and is smaller at large radius. Because the flux density is proportional to $B'^{\frac{p+1}{2}}$ and the emission is roughly uniform in the emitting region at lower energy band (R-band and 5 keV, \cite{1999ApJ...513..679G}), the light-curve peak with a larger $b$ value will be higher for early-time emission (see the first light-curve peak in the two low energy band), while it will be lower for late-time emission (see the second light-curve peak).

The profiles of the PD curves are similar and the PD values show slight difference for three $b$ values (corresponding to $[1b_i]$, $[1c_i]$ and $[1d_i]$ models) at each calculated energy-band. The different $b$ value changes the distribution of the local flux density $f_{\nu}(\theta)$ with the radius, resulting in a change in the contribution of the polarized flux at each radius to the total ones. On one EATS, the local PD (PD($\theta$), which is the PD of the emission from the ring between $\theta$ and $\theta+d\theta$.) at smaller $\theta$ (i.e., large radius) will approach the maximum local PD $\Pi_p$, while PD($\theta$) at larger $\theta$ (i.e., smaller radius) would be smaller. The larger the index $b$ is, the smaller the $f_{\nu}(\theta)$ at the large radius would be, hence the smaller the contribution of the polarized emission at the large radius to the total one from the whole emitting region. Taking the observational frequency of 300 keV as an example, we plot the distribution of $f_{\nu}(\theta)$ and PD($\theta$) with $\theta$ at the corresponding peak time of the light curve (see Figure \ref{reason_with_tobs}). PD($\theta$) of small $\theta$ is close to the local maximum  PD $\Pi_p$, while the $f_{\nu}(\theta)$ contribution at small $\theta$ of the $[1b_i]$ model is higher than that of the $[1c_i]$ model and the $[1d_i]$ model, so the jet PD of the $[1b_i]$ model is higher. Therefore, the values of time-resolved and time-integrated PD with smaller $b$ will be slightly larger at each calculated energy band.

Because the mid-energy photon spectral index $\alpha_2$ only affect the local maximum PD $\Pi_p$, the $\Pi_p$ equals to 0.6 for the $[1b_i]$ model and to 0.69 for the $[1b_i2]$ model. The final jet PD would be obtained by the integration on the EATS. So the PD curves of the $[1b_i]$ and the $[1b_i2]$ models are similar in all the calculated energy bands. Therefore, the mid-energy photon  spectral index $\alpha_2$ show slight impacts on the polarizations.

There are five models involving different variation patterns of $\gamma_{ch}$ in \cite{2018ApJ...869..100U}, as mentioned in Section 2, only $i$ and $m$ models are discussed here. For the $i$ ($[1b_i]$ and $[2b_i]$, with a hard-to-soft $E_p$ evolution mode) and $m$ ($[1b_m]$ and $[2b_m]$, with an intensity-tracking mode) models, the difference of the light curves for the two models becomes obvious in higher energy bands (i.e., 300 keV and 1 MeV). The only difference of the two models is their $E_p$ evolution mode (i.e., the variations of $\gamma_{ch}$ with the radius). And $\gamma_{ch}$ is related to the critical frequency $\nu_{min}$, hence the prominent light-curve difference of the two models would be at the energy band above $\nu_{min}$, corresponding to 300 keV and 1 MeV here.

The PD curves and PD spectra of the $i$ and $m$ models had been carefully studied in \cite{2020ApJ...892..141L}. In their paper, they concluded that PD curves of  the two models would in general decay with time and are weakly correlated with the $E_p$ patterns. The conclusions are not changed and it is analysed more meticulously here. At higher energy bands, if $h\nu>E_p$, then the spectral index is $\beta$. If $h\nu<E_p$, then the spectral index is $\alpha_2$. The local PD $\Pi_p$ will be larger for a larger spectral index. So for the same observational frequency, the PD of the jet emission with a larger $E_p$ would be smaller in general. Here, the $E_p$ value of the two models at the times of their light-curve peak are both around 300 keV and the $E_p$ value will be larger for the $i$ model before its light-curve peak time, hence  compared with the $m$ model, the jet PD will be smaller for $i$ model at early stage.

At early stage (before roughly 0.9 s at which $\theta$ equals to 0 at $r=r_0$), compared with the $[1b_i]$ model (a constant-velocity jet), the bulk Lorentz factor of the $[2b_i]$ model (an accelerating jet shell) is smaller (see our setting of the parameters). It will take a longer time for the $[2b_i]$ model to reach the time of the light-curve peak, hence the first light-curve peak at lower energy bands (i.e., R-band and 5 keV here) and the only light-curve peak at higher energy bands (i.e., 300 keV and 1 MeV here) of the $[2b_i]$ model will be later than the corresponding ones for the $[1b_i]$ model. While the second light-curve peak at lower energy bands of the $[2b_i]$ model will be earlier than the corresponding ones for the $[1b_i]$ model. It is because that when $t_{obs}>0.9$ s the bulk Lorentz factor of the $[2b_i]$ model will be larger than that of the $[1b_i]$ model. The same is for the $[1b_m]$ and $[2b_m]$ models.

We use $[1b_i]$ and $[2b_i]$ models as examples in the following analysis to discuss the effects of the two MFCs. The PD curves of the $[2b_i]$ model (with an aligned field in the emitting region) at lower energy bands (R-band and 5 keV) are similar to that of the $[1b_i]$ model (with a toroidal field in the radiation region). However, its PD evolution trend at higher energy bands (300 keV and 1 MeV) shows obvious differences from that of the $[1b_i]$ model. At early stage, because the jet of the $[1b_i]$ model moves faster than the $[2b_i]$ model, the radiation region for the $[1b_i]$ model will be larger at the same observational time. The field lines becomes less syntropic in the radiation region of the $[1b_i]$ model, while it is always syntropic for an aligned field of the $[2b_i]$ model. Hence, PDs of the $[2b_i]$ model is larger than that of the $[1b_i]$ model.

\begin{figure*}
\centering
\includegraphics[scale=0.32]{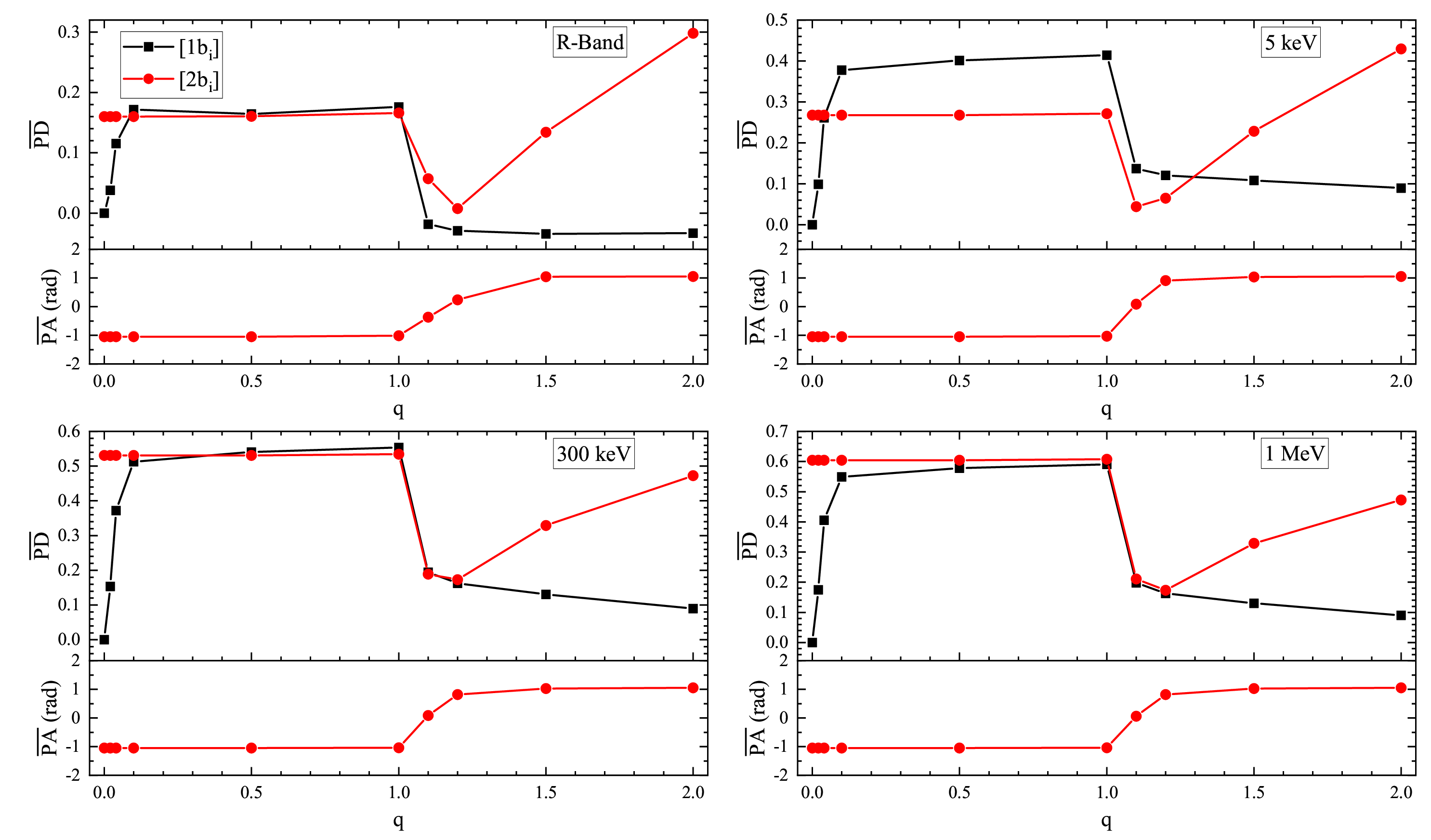}
\caption{The variations of the time-integrated PD and PA with $q=\theta_V/\theta_j$. Upper-left, upper-right, lower-left, and lower-right panels correspond to the observational energy bands of R-band, 5 keV, 300 keV, and 1 MeV, respectively. The black and red lines correspond to the models of $[1b_i]$ and $[2b_i]$, respectively.}
\label{PD_q}
\end{figure*}

Because the decay index of the magnetic field, the mid-energy photon spectral index, and the $\gamma_{ch}$ patten all have slight effect on the jet polarizations, in the following we compare the polarizations of the $[1b_i]$ and $[2b_i]$ models in four energy bands in detail. The variations of the time-integrated PD with  $q$ are shown in Figure \ref{PD_q}, and the corresponding light curves and polarization curves are shown in Figures \ref{lc_q_toroidal} and \ref{lc_q_aligned} in the Appendix \ref{appendix}. And $q$ is the ratio of $\theta_V$ to $\theta_j$. The typical predicted time-integrated PDs will increse with the observational energy band (see Figure \ref{lc_Uhmmodel}). It is roughly $17\%$ at optical R-band, roughly $30\%$ at X-rays 5 keV, roughly $50\%$ at 300 keV, and roughly $55\%$ at 1 MeV.

Because an aligned field with a roughly unchanged asymmetry when $q=\theta_V/\theta_j\leq1$ (on-axis observation) resides in the emission region of the $[2b_i]$ model, its time-integrated PDs when $q\leq1$ stay as constants. While for the $[1b_i]$ model, with the increase of $q$ When $q\leq1/(\Gamma_0\theta_j)$, the field in the main radiation region (i.e., within $1/\Gamma_0$) becomes more and more syntropy. And when When $1/(\Gamma_0\theta_j)<q<1+1/(\Gamma_0\theta_j)$, the field within $1/\Gamma_0$ keeps roughly syntropy. Hence the time-integrated PD values of $[1b_i]$ model will increase with $q$ when $q\leq1/(\Gamma_0\theta_j)$ and then be a roughly constant when $1/(\Gamma_0\theta_j)<q<1+1/(\Gamma_0\theta_j)$. The results here are consistent with that in \cite{SL2024} and \cite{Toma2009}. For off-axis observation ($q=\theta_V/\theta_j>1$), the variation trends of the time-integrated PD of the two models are different. The time-integrated PD of the $[1b_i]$ model would decrease with $q$ in general in the four energy bands, while after the steep decay phase the time-integrated PD of the $[2b_i]$ model will increase with $q$.

\begin{figure*}
\centering
\includegraphics[scale=0.32]{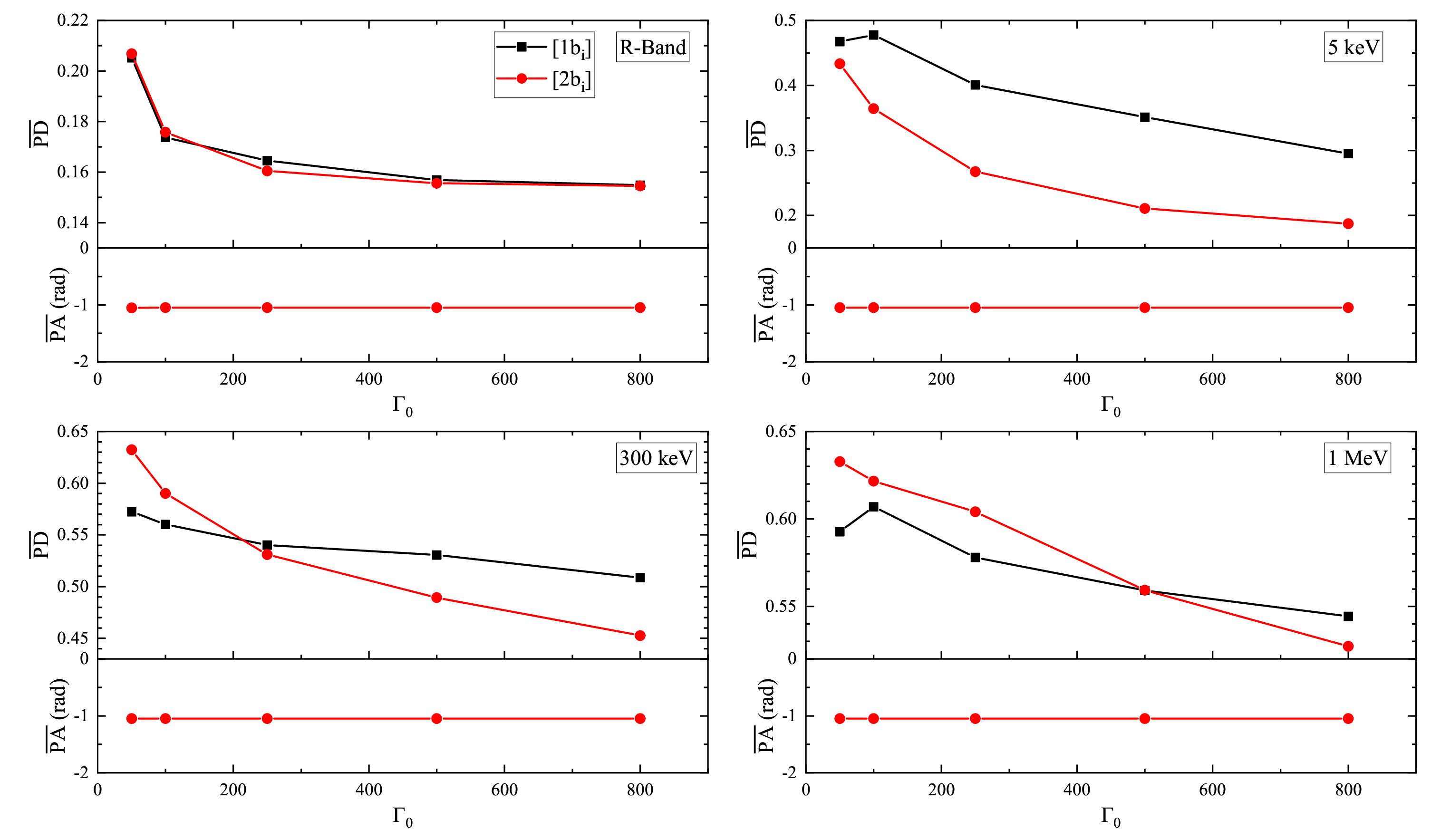}
\caption{The variations of the time-integrated PD and PA with $\Gamma_0$. Upper-left, upper-right, lower-left, and lower-right panels correspond to the observational energy bands of R-band, 5 keV, 300 keV, and 1 MeV, respectively. The black and red lines correspond to the models of $[1b_i]$ and $[2b_i]$, respectively.}
\label{PD_g0}
\end{figure*}

\begin{figure*}
\centering
\includegraphics[scale=0.32]{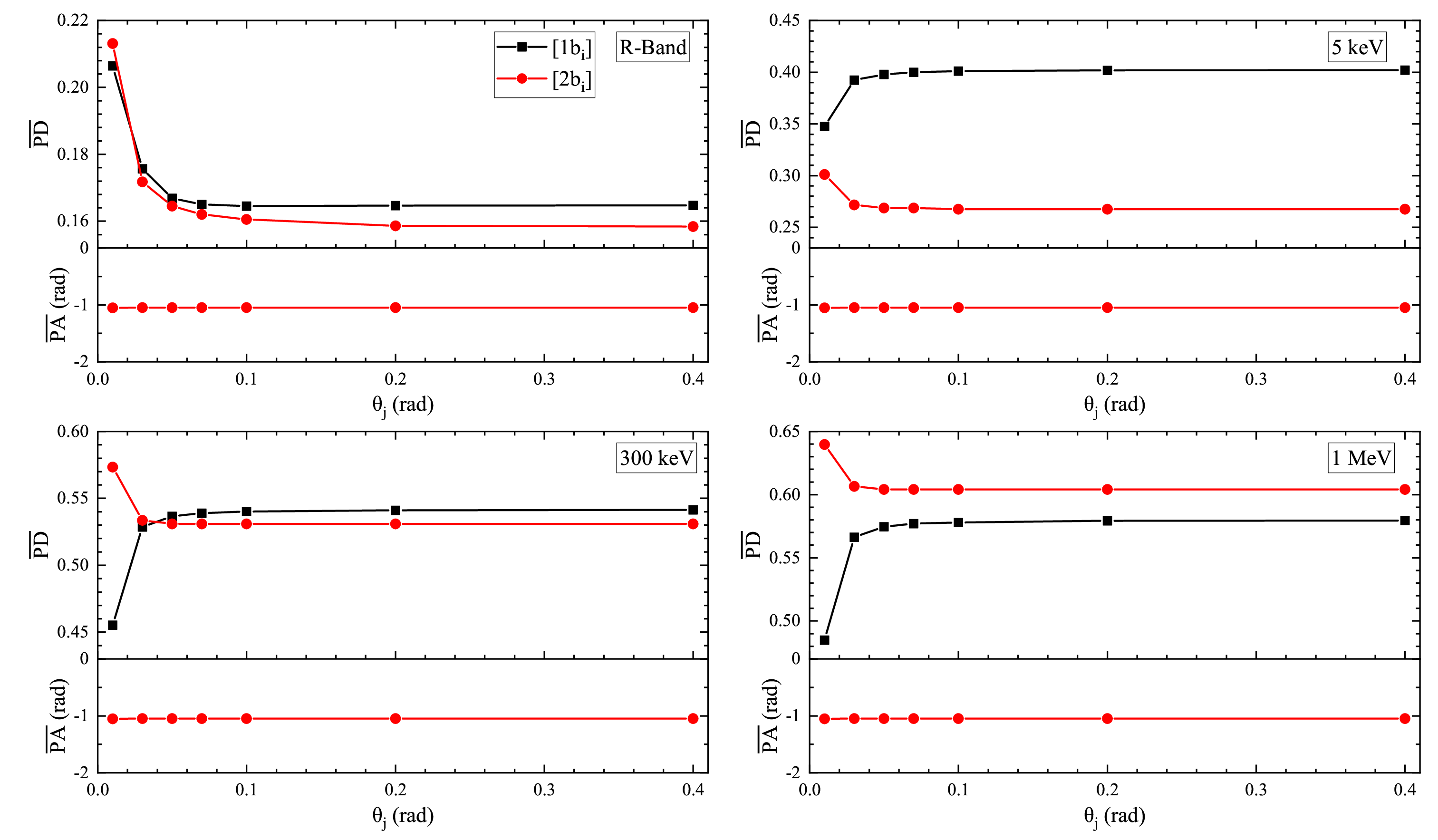}
\caption{The variations of the time-integrated PD and PA with $\theta_j$. Upper-left, upper-right, lower-left, and lower-right panels correspond to the observational energy bands of R-band, 5 keV, 300 keV, and 1 MeV, respectively. The black and red lines correspond to the models of $[1b_i]$ and $[2b_i]$, respectively.}
\label{PD_thetaj}
\end{figure*}

The variations of the time-integrated PD with both $\Gamma_0$ and $\theta_j$ are also investigated using $[1b_i]$ and $[2b_i]$ models. The results are shown in Figures \ref{PD_g0} and \ref{PD_thetaj}, and the corresponding results of the light curves and polarization curves are shown in Figures \ref{lc_g0} and \ref{lc_thetaj} in the Appendix \ref{appendix}. When $\Gamma_0$ increase, the $1/\Gamma$ cone becomes smaller, so the radiation from outside of the $1/\Gamma$ cone will be larger. Therefore, the time-integrated PD would decrease with $\Gamma_0$ in general. The time-integrated PD of the $[2b_i]$ model decreases with $\theta_j$ in all four energy band, while it will decrease in optical R-band with $\theta_j$ and will increase with $\theta_j$ in 5 keV, 300 keV and 1 MeV for the $[1b_i]$ model.

\section{Conclusions and Discussion}\label{sec:Conclusions and Discussion}

In this paper, we use a three-segment power-law spectrum to reconstruct the Stokes parameters of the jet emission in GRB prompt phase. Then the multi-band light curves, polarization curves and the time-integrated polarizations from optical R-band to MeV gamma-rays are studied in detail. The synchrotron emission is assumed here. The MFC in the radiation region is assumed to be large-scale ordered, hence the predicted PD would be the upper limit.

Because of a decaying maximum flux density and the evolution of the minimum critical frequency, the light curves at the lower energy bands (R-band \& 5 keV) have two peaks, while there is only one at higher energy bands (300 keV and 1 MeV).  A decaying proportion of the emission from the region within $1/\Gamma$ cone will lead to a decrease of the jet PD and vise versa. So the profiles of the PD curves are positively correlated with that of the $\tilde{f}$ parameter (except $\tilde{f}=0$ and $\infty$). Due to the evolution of the $\tilde{f}$ parameter, some PD curves decreases at early times, then rises shallowly, and finally continues to decay at late stages. PA is roughly a constant during the main radiation episode at the optical and gamma-ray bands for the synchrotron models here, while it evolves randomly for the photosphere model \citep{2022ApJ...926..104P}. Therefore, the synchrotron models and the photosphere model could be distinguished by the time-resolved PA observation.

The product value of $\Gamma_0$ and $\theta_j$ has a great influence on the polarization. For $\Gamma_0\theta_j>1$, there will be a long-lasting high PD phase, which would even extend to out of the $T_{95}$. The reason for this long lasting high PD phase is dynamical. The general profile of the PD curves will not be affected by the concrete peak-energy evolution mode. However, there might be a small PD peak at early stage for the hard-to-soft mode and it is almost unlikely for the intensity-tracking mode \citep{2021ApJ...909..184L}. The time-integrated PD of the synchrotron model would decrease with $\Gamma_0$ due to the increase of the radiation outside $1/\Gamma$ cone, while it  will increase with the observational energy band. The profile of the $q-\bar{PD}$ curve is similar for the same MFC in the radiation region and is independent of the observational energy band.

At the optical R-band, the time-integrated PD would be roughly $17\%$ in the synchrotron model studied here, while the maximum value of the time-resolved one can reach $\sim50\%$. The predicted time-integrated PD here is similar to that of the photosphere model \citep{2022ApJ...926..104P}. Up till now, there is only one polarization detection in optical band by the MASTER net telescopes during the prompt phase of GRB 160625B \citep{Troja2017}. The detected time-resolved PD at the decay phase of the optical light curve is $\sim5\%$. Then with the flatten of the optical light curve, it increases to $\sim8\%$. Theoretically, at the decay phase of the optical light curve, PD also decreases sharply with time down to zero. Both the increase of the observed PD and the flattened light curve in optical band indicate a new radiation component. There are two possible reasons for the detected moderate $\sim8\%$ PD. One is that the magnetic field in the radiation region is mixed as mentioned in \citep{Troja2017}. The other is because of the mixture of the two offsetting polarized radiation components. In optical band, the Very Large Telescope (VLT) and Liverpool Telescope (LT) could also do polarization detections. The predicted typical $T_{90}$ of the prompt optical radiation is rougly 100 s and the flux density is around 50 $\mu$Jy, which could be detected by these facilities.

At X-ray band, the only burst with prompt X-ray polarization detection is GRB 221009A and the observed upper limit of the PD ranges from $\sim55\%$ to $\sim82\%$ with $99\%$ confidence level by the Imaging X-ray Polarimetry Explorer (IXPE) \citep{2022JATIS...8b6002W,Negro2023}. The predicted time-resolved PD can be as high as $60\%$ and the time-integrated one is roughly around $(30-40)\%$ here. Our results here are consistent with the observation. However, because of harsh conditions such observation are quite rare. In the near future, the Low-energy Polarimetry Detector (LPD) on board POLAR-2 \citep{POLAR2} would work. The predicted fluence at X-ray band is about $4\times10^{-7}$ erg cm$^{-2}$, which might be deficient for polarization analysis. However, for the bright bursts the fluences might reach $4\times10^{-5}$ erg cm$^{-2}$, which would be sufficient for polarization analysis.

In gamma-ray band, both the time-integrated and time-resolved PD values are predicted to be roughly zero for the photosphere model \citep{2022ApJ...926..104P,2018ApJ...856..145L}. While the upper limit of the time-integrated one is around $50\%$ for the synchrotron models here. So the models and hence their radiation mechanism can be tested by the PD detection in gamma-ray band. Up till now, we have about 40 bursts with polarization detection in gamma-ray band  \citep{Yonetoku2011,Yonetoku2012,Zhang2019,Kole2020,Chattopadhyay2022}. However, due to the large observational errors the average value of the observed PD can not be determined accurately. The high-energy Polarimetry Detector (HPD) on board POLAR-2 \citep{POLAR2} could distinguish between the $0\%$ polarization and the $10\%$ polarizaiton \footnote{The observed PD values of the bursts by POLAR are concentrated around $10\%$. However, the $0\%$ polarization can not be ruled out \citep{Zhang2019,Kole2020}.}. Therefore, the detection results of HPD will tell us for most bursts which model will work. The Compton Spectrometer and Imager (COSI), which was planned to launch in 2027, could also detect GRB prompt polarizations \citep{2023arXiv230812362T}. In a two-years mission, it will detect about 40 bursts with polarization observations and the sample would be ampliative. Exciting new phenomenons might be found.

\begin{acknowledgements}
We thank the anonymous referee for helpful comments that dramatically improved our paper. This work is supported by the National Natural Science Foundation of China (grant No. 11903014). M.X.L also would like to appreciate the financial support from Jilin University. 
\end{acknowledgements}

\appendix
\restartappendixnumbering

\section{The light curves and polarization curves with different parameters}\label{appendix}

The impact of the bulk Lorentz factor $\Gamma_0$, jet opening angle $\theta_j$, observational angle $\theta_V$, and MFCs on the light curves and polarizations are investigated. The model parameters follow Table. \ref{Model Parameters}. The orientation of the aligned magnetic field is set to be $\delta=\pi/6$. Other parameters are fixed as follows: $B_0=30$ G, $r_{on}=10^{14}$ cm, $r_{off}=3\times10^{16}$ cm, $\gamma_{ch}^i=5\times10^4$, and $r_0=10^{15}$ cm \citep{2018ApJ...869..100U}.

\begin{figure*}
\centering
\includegraphics[scale=0.32]{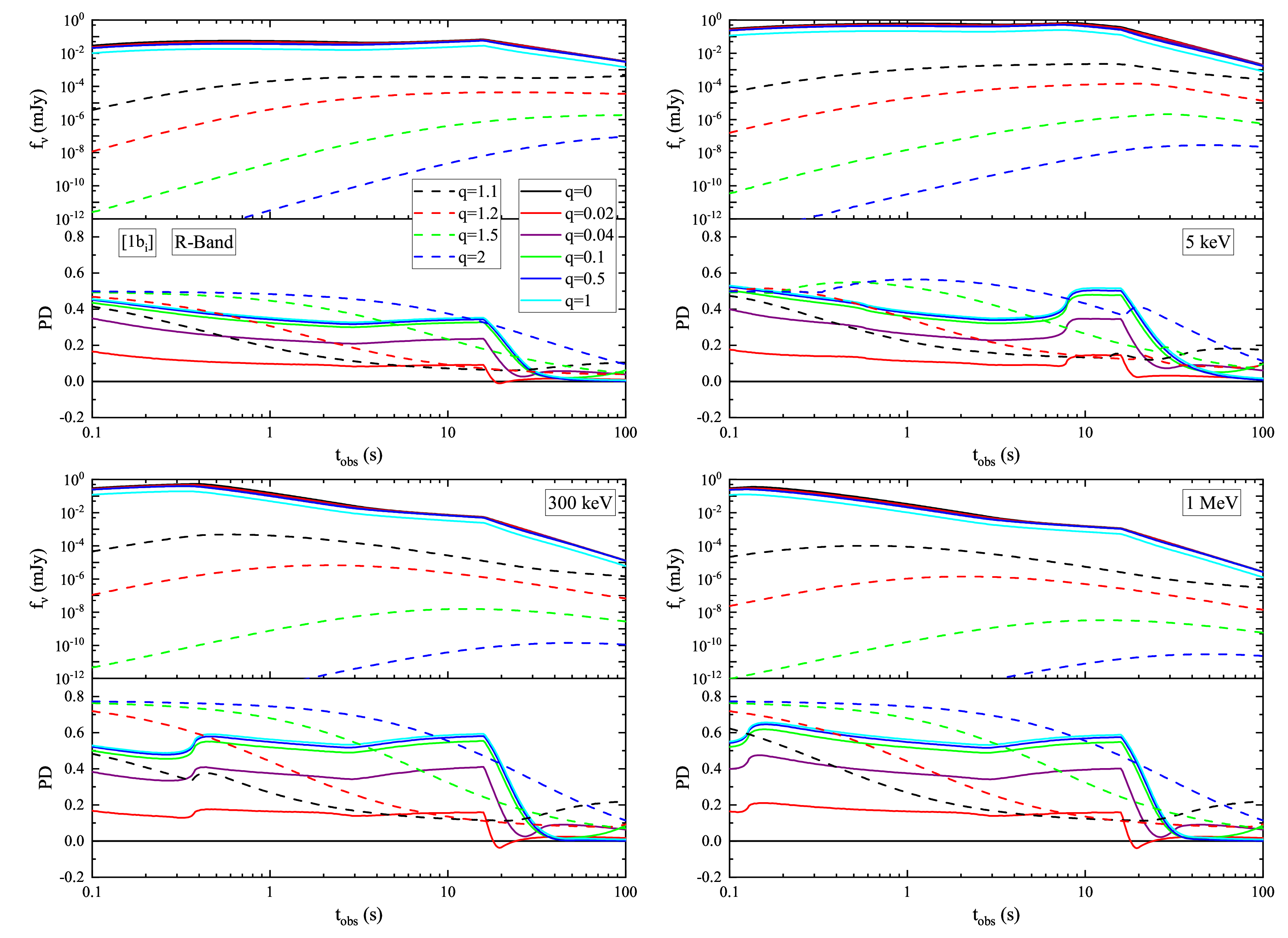}
\caption{Light curves and polarization evolutions of the $[1b_i]$ model for various $q=\theta_V/\theta_j$ values. Upper-left, upper-right, lower-left, and lower-right panels correspond to the observational energy bands of R-band, 5 keV, 300 keV, and 1 MeV, respectively. In each panel, top and bottom panels show the light and PD curves, respectively. The black, red, purple, green, blue and cyan solid lines correspond to $q=0$, 0.02, 0.04, 0.1, 0.5 and 1, respectively. The black, red, green and blue dashed lines correspond to $q=1.1$, 1.2, 1.5 and 2.0, respectively.}
\label{lc_q_toroidal}
\end{figure*}

\begin{figure*}
\centering
\includegraphics[scale=0.32]{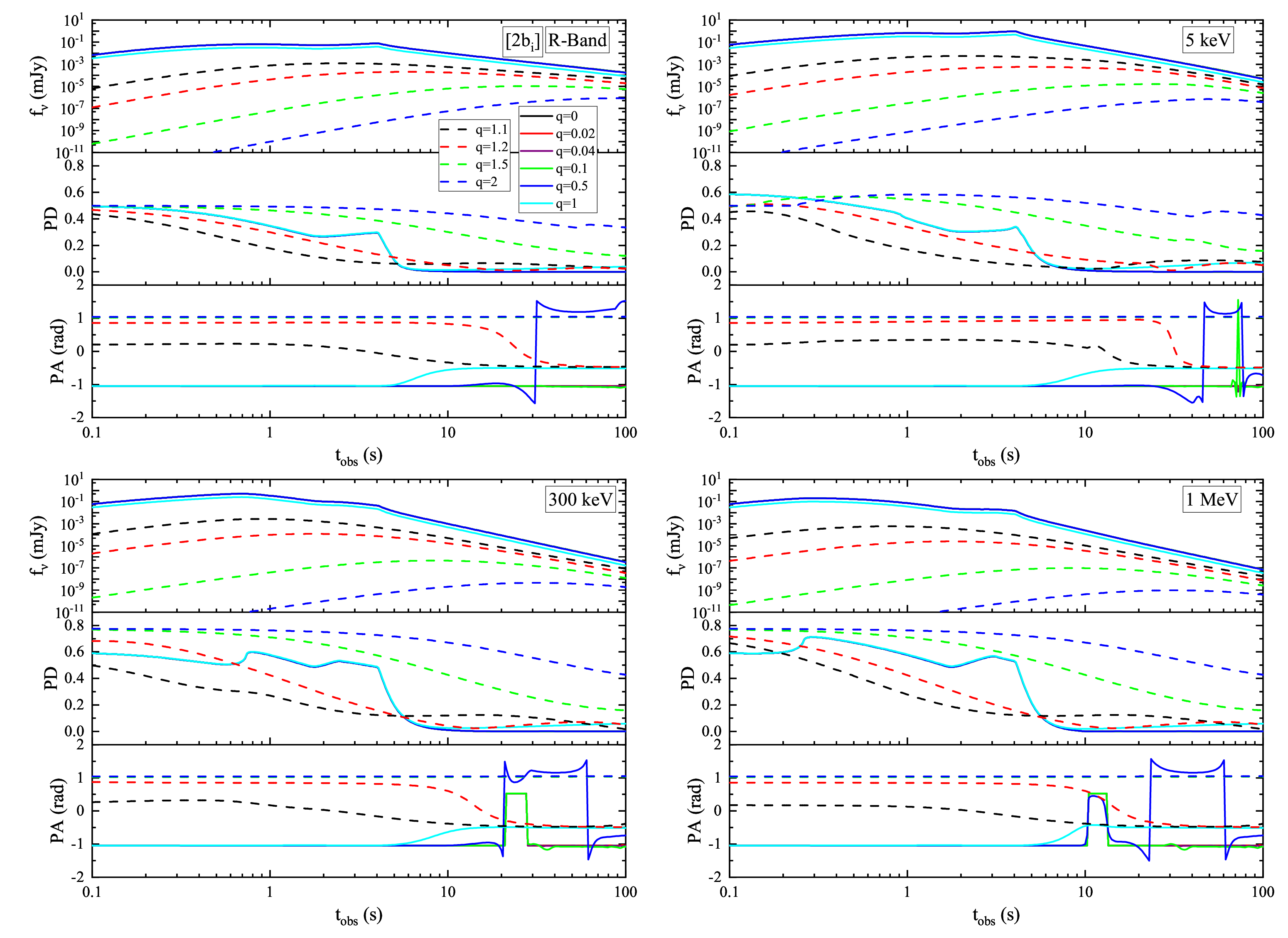}
\caption{Light curves and polarization evolutions of the $[2b_i]$ model for various $q=\theta_V/\theta_j$ values. Upper-left, upper-right, lower-left, and lower-right panels correspond to the observational energy bands of R-band, 5 keV, 300 keV, and 1 MeV, respectively. In each panel, top, middle, and bottom panels show the light, PD and PA curves, respectively. The black, red, purple, green, blue and cyan solid lines correspond to $q=0$, 0.02, 0.04, 0.1, 0.5 and 1, respectively. The black, red, green and blue dashed lines correspond to $q=1.1$, 1.2, 1.5 and 2.0, respectively.}
\label{lc_q_aligned}
\end{figure*}

We fixed $\Gamma_0=250$ and $\theta_j=0.1$ rad, and varied $\theta_V$ to calculate the light and polarization curves of the $[1b_i]$ and $[2b_i]$ models. The numerical results are shown in Figures \ref{lc_q_toroidal} and \ref{lc_q_aligned}. The  PD curves of the $[2b_i]$ model with different $q$ are almost coincide with each other  for on-axis observations ($q=\theta_V/\theta_j\leq1$). While for the $[1b_i]$ model, the PD values will increase with $q$ when $q\leq1/(\Gamma_0\theta_j)$ and then be roughly consistent when $1/(\Gamma_0\theta_j)<q<1+1/(\Gamma_0\theta_j)$. There are many abrupt PA rotations or oscillations at late evolution stage. The time and the number of the rotation or the oscillation depend on the computational accuracy when the values of the Stokes parameters becomes small at late stage.

\begin{figure*}
\centering
\includegraphics[scale=0.32]{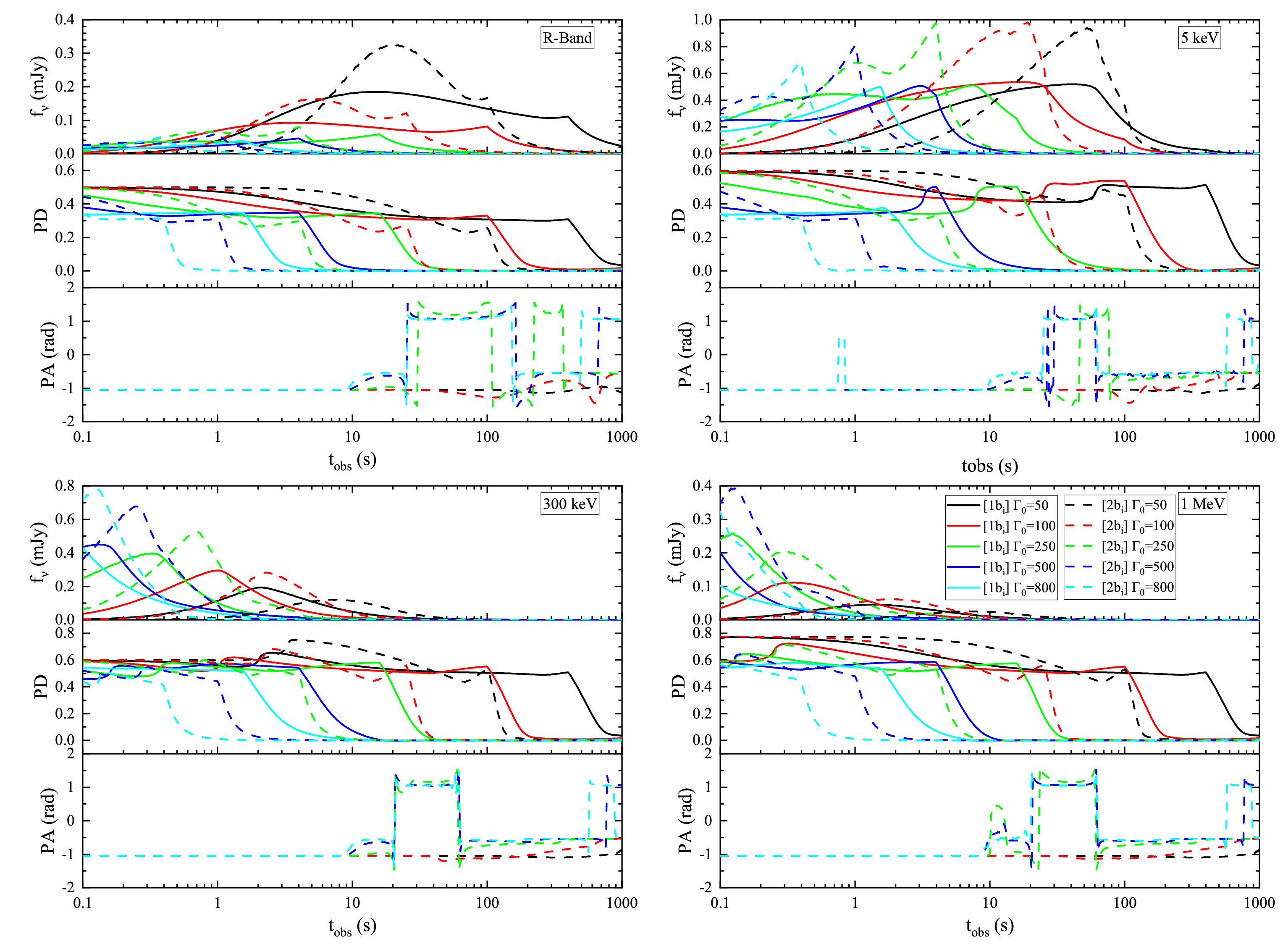}
\caption{Light curves and polarization evolutions for various bulk Lorentz factor $\Gamma_0$. Upper-left, upper-right, lower-left, and lower-right panels correspond to the observational energy bands of R-band, 5 keV, 300 keV, and 1 MeV, respectively. In each panel, top, middle, and bottom panels show the light, PD and PA curves, respectively. The solid and dashed lines correspond to models of $[1b_i]$ and $[2b_i]$, respectively. The black, red, green, blue, and cyan lines correspond to $\Gamma_0=50$, 100, 250, 500 and 800, respectively.}
\label{lc_g0}
\end{figure*}

We fixed $\theta_j=0.1$ rad and $\theta_V=0.05$ rad, and varied $\Gamma_0$ to calculate  the light and polarization curves of the $[1b_i]$ and $[2b_i]$ models. The numerical results are shown in Figure \ref{lc_g0}. The peak of the light curve will move toward large time with a decrease of $\Gamma_0$. And independent of the observational energy band, the lasting timescales of the PD plateau phase will decrease as the $\Gamma_0$ increases.

\begin{figure*}
\centering
\includegraphics[scale=0.32]{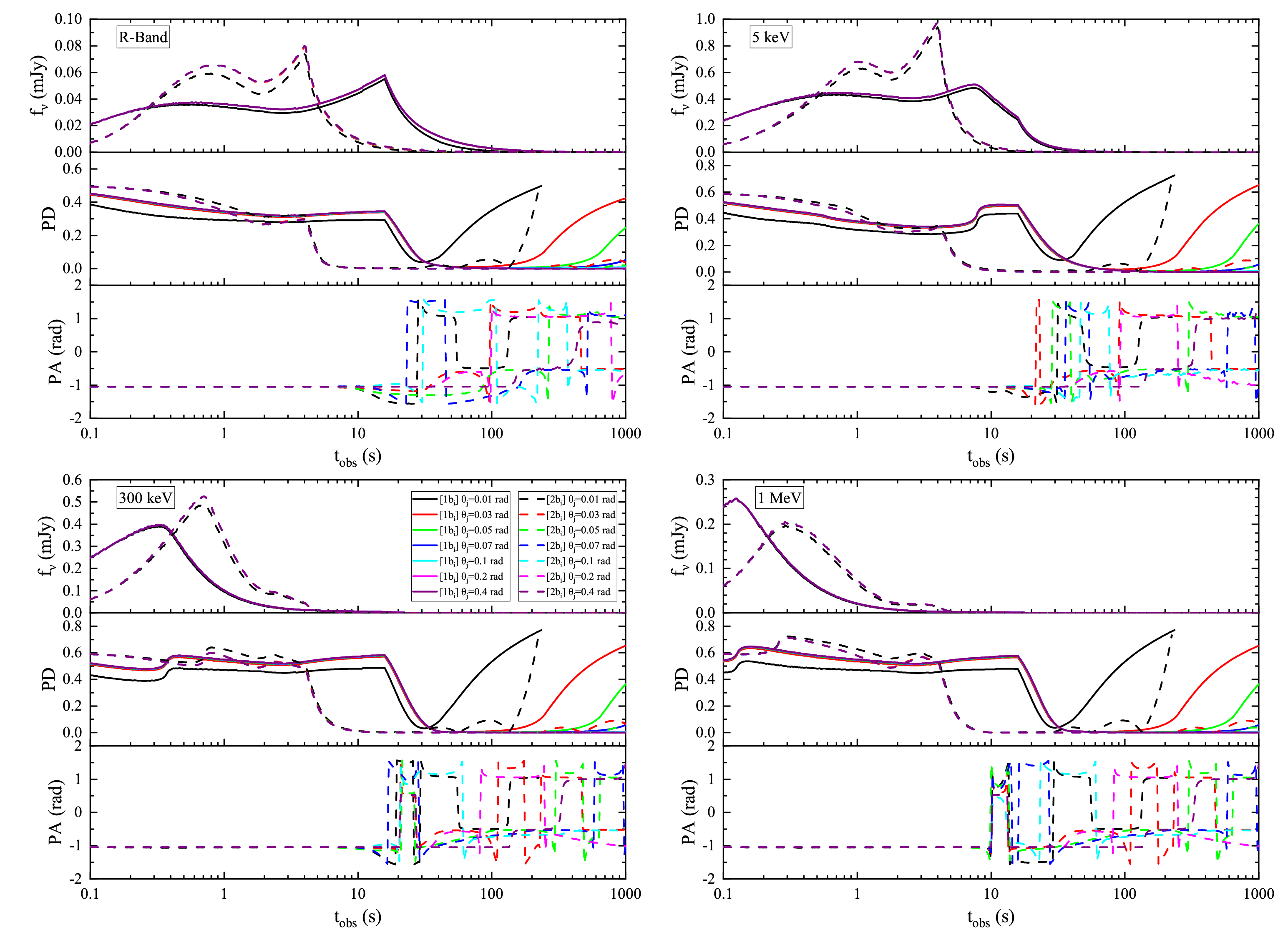}
\caption{Light curves and polarization evolutions for various jet opening angle $\theta_j$. Upper-left, upper-right, lower-left, and lower-right panels correspond to the observational energy bands of R-band, 5 keV, 300 keV, and 1 MeV, respectively. In each panel, top, middle, and bottom panels show the light, PD and PA curves, respectively. The solid and dashed lines correspond to models of $[1b_i]$ and $[2b_i]$, respectively. The black, red, green, blue, cyan, pink and purple lines correspond to $\theta_j=0.01$, 0.03, 0.05, 0.07, 0.1, 0.2 and 0.4 rad, respectively.}
\label{lc_thetaj}
\end{figure*}

We fixed $\Gamma_0=250$ rad and $q=0.5$, and varied $\theta_j$ to study the light and polarization curves of the $[1b_i]$ and $[2b_i]$ models. The numerical results are shown in Figure \ref{lc_thetaj}. It is found that the PD curves with $\theta_j=0.01$ and 0.03 rad will rise at the late stage before the flux density disappears. We then extended the time range to 10000 s, and found the PD curves with other $\theta_j$ values have the similar fast rising PD profile at late stage. This is because the field lines in the observed radiation region becomes more syntropic with the shrinking of its area at late stage.

\listofchanges

\bibliography{ms_arxiv}

\end{document}